\documentclass[letterpaper,table]{zadpreprints_adj}
\pdfoutput=1

\usepackage{ragged2e}
\sectionfont{\RaggedRight}

\newcommand{\eightpt}{\fontsize{7}{5}\selectfont}
\renewcommand{\footnotesize}{\eightpt}

\makeatletter 
\renewcommand{\@makefntext}[1]{%
  \setlength{\parindent}{0pt}%
  \begin{list}{}{\setlength{\labelwidth}{2mm}
    \setlength{\leftmargin}{40pt}%
    \setlength{\labelsep}{1pt}%
    \setlength{\itemsep}{0pt}%
    \setlength{\parsep}{0pt}%
    \setlength{\topsep}{0pt}%
    \color{black}\footnotesize}%
  \item[\@thefnmark\hfil]#1
  \end{list}%
}
\makeatother 

\graphicspath{ArXiV/}
\DeclareUnicodeCharacter{2212}{-}

\papertype{Statistical Theory}

\title{\Large Technical Issues in the Interpretation of S-values and Their Relation \\to Other Information Measures}

\citearticle{Greenland, S. \& Rafi, Z. Technical Issues in the Interpretation of S-values and Their Relation to Other Information Measures. \textit{arXiv:1909.08579 [stat.TH]} (2020).}

\author[1]{Sander Greenland \protect\orcidicon{0000-0003-4364-3279}}
\author[2]{Zad Rafi \protect\orcidicon{0000-0003-1545-8199}}

\affil[1]{Department of Epidemiology and Department of Statistics, 
University of California, Los Angeles, CA}
\affil[2]{Department of Population Health, NYU Langone, New York, NY}

\corraddress{Sander Greenland}
\corrcontact{
	$^{1}$Email: lesdomes@ucla.edu \\$^{2}$Twitter: \href{https://twitter.com/Lester_Domes}{@Lester\_Domes} \\
	$^{2}$Email: zad@lesslikely.com \\$^{1}$Twitter: \href{https://twitter.com/DailyZad}{@DailyZad}}

\hypersetup{
pdftitle={Technical Issues in the Interpretation of S-values and Their Relation to Other Information Measures.},
pdfsubject={Statistical Theory},
pdfauthor={Sander Greenland and Zad Rafi},
pdfdate={2020-09-20},
pdfproducer=pdfLaTeX,
pdfcopyright={Copyright (C) 2020,  The Author(s)},
pdfkeywords={Information Statistics, Statistical Theory, Bayesian Statistics, Statistical Models, Data Interpretation, Hypothesis Tests, P-values, S-values, Statistical Significance},
pdflang={en},
pdfmetalang={en}
}

\begin{document}

\maketitle

\begin{abstract}
\vspace{0.2cm}
An extended technical discussion of \textit{S}-values and unconditional information can be found in Greenland \cite{greenlandValidPvaluesBehave2019} and Greenland \& Rafi \cite{greenlandAidScientificInference2020}. Here we briefly cover several technical topics mentioned in our \href{https://doi.org/10.1186/s12874-020-01105-9}{main paper} \cite{rafiSemanticCognitiveTools2020}: Different units for (scaling of) the \textit{S}-value besides base-2 logs (bits); the importance of uniformity (validity) of the \textit{P}-value for interpretation of the \textit{S}-value; the combination of the \textit{S}-value across studies; and the relation of the \textit{S}-value to other measures of statistical information about a test hypothesis or model.

\vspace{-0.1cm}
\vspace{0.2cm}
\end{abstract}
\vspace{-.5cm}

\begin{adjustwidth*}{60pt}{60pt}
\section{Background}
\label{sec:Background}
{\fontsize{8.5}{13}\selectfont{
An extended technical discussion of \textit{S}-values and unconditional information can be found in Greenland \cite{greenlandValidPvaluesBehave2019} and Greenland \& Rafi \cite{greenlandAidScientificInference2020}. Here we briefly cover several technical topics mentioned in our \href{https://doi.org/10.1186/s12874-020-01105-9}{main paper} \cite{rafiSemanticCognitiveTools2020}: Different units for (scaling of) the \textit{S}-value besides base-2 logs (bits); the importance of uniformity (validity) of the \textit{P}-value for interpretation of the \textit{S}-value; the combination of the \textit{S}-value across studies; and the relation of the \textit{S}-value to other measures of statistical information about a test hypothesis or model.}}

\section{Units for the \textit{S}-value}
{\fontsize{8.5}{13}\selectfont{
Other units for measuring information other than bits arise from different choices for the base of the logarithms. For example, using natural (base-e) logs, the \textit{S}-value becomes $s_{e}$ = $−\ln(p)$ = $-\log_{2}(p)\ln(2)$ whose units are called “nats,” while using common (base-10) logs the \textit{S}-value becomes $s_{10}$ = $-\log_{10}(p)$ = $-\log_{2}(p)\log_{10}(2)$ whose units are called hartleys, bans, or dits (decimal digits). The ratio of one dit of information to one bit of information is $\log_{2}(10)$ = 3.22 which is similar to the ratio of meters to feet, 3.28. Just as the choice of meters vs. feet does not affect the concepts and methods surrounding length measurement, so choice of dits vs. bits does not affect any of the concepts or methods of information measurement. Bits are most commonly used in communications engineering because the fundamental physical components in electronic information storage are binary and thus their information capacity is one bit. Natural logs are however more mathematically convenient and thus more common in statistical theory (see below), although base-10 logs are also seen.}}

\newpage
\noindent\textcolor{gray!50}{\rule[-0.05ex]{\linewidth}{0.05pt}}
\section[Uniformity of the \textit{P}-value and the information in the \textit{S}-value]{\texorpdfstring{Uniformity of the \textit{P}-value \\and the information in the \textit{S}-value}{Uniformity of the \textit{P}-value and the information in the \textit{S}-value}}

 {\fontsize{9}{13}\selectfont{
The decision rule “reject \textbf{H} if $p\leq$ $\alpha$” will reject \textbf{H} 100$\alpha$\% of the time under sampling from a model \textbf{M} obeying \textbf{H} (i.e., the Type-1 error rate of the test will be $\alpha$) provided the random variable \textit{P} corresponding to \textit{p} is valid (uniform under the model \textbf{M} used to compute it), but not necessarily otherwise \cite{kuffnerWhyArePvalues2019}. This is one reason why frequentist writers reject invalid \textit{P}-values (such as posterior predictive \textit{P}-values, which highly concentrate around 0.50) and devote considerable technical coverage to uniform \textit{P}-values \cite{kuffnerWhyArePvalues2019,robinsAsymptoticDistributionValues2000,bayarriValuesCompositeNull2000}. A valid \textit{P}-value (“U-value”) translates into an exponentially distributed \textit{S}-value with a mean of 1 nat or $\log_{2}(e)=1.443$ bits where $e$ is the base of the natural logs. \\ \indent
Uniformity is also central to the “refutational information” interpretation of the \textit{S}-value used here, for it is necessary to ensure that the \textit{P}-value \textit{p} from which \textit{s} is derived is in fact the percentile of the observed value of the test statistic in the distribution of the statistic under \textbf{M}, thus making small \textit{p} surprising under \textbf{M} and making \textit{s} the corresponding degree of surprise. Because posterior predictive \textit{P}-values do not translate into sampling percentiles of the statistic under the hypothesis (in fact, they are pulled toward 0.5 from the correct percentiles) \cite{robinsAsymptoticDistributionValues2000,bayarriValuesCompositeNull2000}, the resulting negative log does not measure surprisal at the statistic given \textbf{M}, and so is not a valid \textit{S}-value in our terms. \\ \indent
For simplicity, we have assumed that at least an approximately valid \textit{P}-value can be derived for testing \textbf{M}. This is so in typical regression analyses in health and medical sciences, but not always. To deal with exceptions, \textit{P}-values are often said to be “conservatively valid” for testing \textbf{M} when under \textbf{M} they stochastically dominate a unit-uniform distribution, i.e., under \textbf{M} the probability that \textit{P} exceeds a given \textit{p} is at least \textit{p}, and for some \textit{p} exceeds \textit{p}. Typical exact \textit{P}-values from discrete data are conservatively valid but approach uniformity with increasing sample size. In cases for which we can only deduce that \textit{P} is conservatively valid (as when its observed value \textit{p} is an upper bound rather than a direct tail probability), we would interpret its corresponding \textit{S}-value as conservatively valid in the sense of representing the minimum information against \textbf{M} supplied by the test. \\ \indent
The coin-toss interpretation we have used to physically gauge this information assumes that the only alternative to fairness is in the direction of loading for heads. The \textit{S}-value it produces thus corresponds to a \textit{P}-value for the 1-sided hypothesis Pr(heads)$\leq \nicefrac{1}{2}$; nonetheless, this interpretation applies even if the original observed \textit{P}-value \textit{p} was 2-sided. This translation from a 2-sided \textit{P}-value to a 1-sided \textit{S}-value parallels the transformation of \textit{P}-values into 1-sided sigmas (\textit{Z}-scores) in physics, in which for example a \textit{P}-value of 0.05 from a two-sided test would become a sigma of 1.645, the upper one-sided 5\% cutoff for a standard-normal deviate \cite{cousinsJeffreysLindleyParadox2017}.}}

\section{Combination of \textit{S}-values across studies}
{\fontsize{9}{13}\selectfont{
Preprints of our article referred to the \textit{S}-value as “additive” over independent sources, which is incorrect insofar as the combined refutational information from independent tests of the same model \textbf{M} is a subadditive function of the separate \textit{S}-values; only the combination of the latter into a summary test statistic is additive. More precisely, suppose we have K studies, each contributing an independent valid \textit{S}-value $S_{k}$ for \textbf{M}. Under \textbf{M}, each $S_{k}$ has an expected value of 1 nat, which can be viewed as the expected “noise” contribution to the \textit{S}-value (Good \cite{goodSurpriseIndexMultivariate1956,goodCorrectionsSurpriseIndex1957} dealt with this factor by what in our case reduces to subtracting 1 nat from each surprisal, which however creates problematic negative values when \textit{s} < 1 nat; we have instead chosen to follow the subsequent theoretical literature and not do so). The sum $S_{+}$ =  $\sum_{k}S_{k}$ will thus have an expectation of K “noise” nats under \textbf{M}. Furthermore, the distribution of $2S_{+}$ will be $\chi^{2}$ on 2K degrees of freedom \cite{coxChapterPureSignificance1974}; hence the summary \textit{S}-value $S_{\&}$ derived from the sum $S_{+}$ will be the negative log of the \textit{P}-value from comparing $2S_{+}$ to a 2K df $\chi^{2}$ distribution. Under \textbf{M} this summary-$\chi^{2}$ $S_{\&}$ has an expectation of 1 nat and thus will on average be K − 1 nats smaller than $S_{+}$, with even larger discrepancies under violations of \textbf{M} that the test is sensitive to. \\ \indent
More generally, if the test model $\textbf{M}_{k}$ varies across studies, the \textit{S} - summation test just given is valid for testing the conjunction (intersection) hypothesis $\textbf{M}_{\&}$ = $\textbf{M}_{1}\&\cdot\cdot\cdot\&\textbf{M}_{K}$, but is rarely optimal because it makes no use of homogeneity or other relations among the models. In particular, the test remains valid when as usual each study model $\textbf{M}_{k}$ incorporates a shared target or focal hypothesis \textbf{H} across studies (e.g., no association of a given treatment and disease), combined with different sets of background assumptions $\textbf{A}_{k}$ (e.g., due to differences in study designs), so that $\textbf{M}_{k}$ = $\textbf{H}\&\textbf{A}_{k}$ and $\textbf{M}_{\&}=\textbf{H}\&\textbf{A}_{1}\&\cdot\cdot\cdot\&\textbf{A}_{K}= \textbf{H}\&\textbf{A}_{\&}$. When however a homogeneity assumption is correct, this general test will have lower power (sensitivity) for violations of \textbf{H} given the background $\textbf{A}_{\&}$ than the usual summary tests (which use homogeneity in deriving the study-specific contributions to their summary statistics). On the other hand, those usual tests can have lower power if homogeneity is very wrong. In any case, the \textit{S}-values from the two tests can differ considerably due to the additional (and possibly incorrect) homogeneity information used in the usual tests. \\ \indent
Some insight into these results may follow from reviewing the traditional parallel procedure for combining \textit{Z}-scores $Z_{k}$ (e.g., standardized residuals) by squaring and summing them. The resulting sum of squares $\sum_{k}Z_{k}^{2}$ has a K df $\chi^{2}$ distribution with expectation K if no cross-k information is used to compute the $Z_{k}$. More generally however the df are reduced by the number of cross-study sharp constraints used. For example, if \textbf{H} is a hypothesis that a mean difference $\delta_ {k}$ is 0, the test using the assumption that the $\delta_ {k}$ are constant across studies imposes K − 1 constraints ($\delta_{1}=\cdot\cdot\cdot=\delta_{K}$) to derive the $Z_{k}$, and so will have only 1 df instead of K df; this reduction in df produces considerably more power if the assumption is correct, but not necessarily otherwise. See Greenland and Rafi \cite{greenlandAidScientificInference2020} for details and examples.}}

\section[Other measures of statistical information about a test hypothesis or model]{\texorpdfstring{Other measures of statistical \\information about a test hypothesis or model}{Other measures of statistical information about a test hypothesis or model}}
{\fontsize{9}{13}\selectfont{
A common measure for evaluating a hypothesis or model restriction \textbf{H} under background assumptions or unrestricted model \textbf{A} is the maximum-likelihood ratio (MLR), which is the value of the likelihood function at its maximum under \textbf{A} alone, divided by its (restricted) maximum when the test hypothesis \textbf{H} is additionally imposed \cite{cummingsAnalysisIncidenceRates2019,jewellStatisticsEpidemiology2003}. The MLR defined this way is always above 1; it is however sometimes confused with the posterior odds against the tested value \textbf{H} given \textbf{A}, which it equals only under very special (and usually unrealistic) conditions. The MLR does however show the most extreme increase in posterior odds against \textbf{H} that the data could produce given \textbf{A}. The corresponding information measure paralleling the \textit{S}-value is the deviance difference or likelihood-ratio (LR) statistic for \textbf{H} given \textbf{A}, $2\ln(MLR)$, which is itself a test statistic for \textbf{H} given \textbf{A}. The change in the Akaike Information Criterion (without small-sample adjustment) from adding \textbf{H} to the background model is $2\ln(MLR)−2d$ where \textit{d} is the dimension (degrees of freedom) of \textbf{H} \cite{burnhamModelSelectionMultimodel2002,cummingsAnalysisIncidenceRates2019}. \\ \indent
Now consider a sharp constraint hypothesis \textbf{H} with a \textit{P}-value less than $\nicefrac{1}{e}$ = 0.368. Bayarri \& Berger \cite{bayarriQuantifyingSurpriseData1999} and Sellke et al. \cite{sellkeCalibrationValuesTesting2001} show that $b=−e\cdot p\cdot \ln(p)$ = $e\cdot p\cdot s_{e}$ is a sharp lower bound on the \textit{Bayes factor} for \textbf{H} under \textbf{A}, where \textbf{A} now includes strong restrictions on the alternatives to \textbf{H}. (A Bayes factor is the ratio of posterior data probabilities under \textbf{H} and an alternative, given \textbf{A}.) Thus, given \textbf{A} and the data, $b$ is a lower bound on the reduction in odds for \textbf{H} given \textbf{A} in moving from a prior to a posterior, and $\nicefrac{1}{b}$ is an upper bound on the increase in odds against \textbf{H} given \textbf{A}. Simple numeric examples show that the latter bound is much lower than the MLR. The strength of the restrictions added to \textbf{A} is indicated for example by the fact that for \textit{p} = 0.05 the MLR in \href{https://bmcmedresmethodol.biomedcentral.com/articles/10.1186/s12874-020-01105-9/tables/1}{Table 1} of our \href{https://doi.org/10.1186/s12874-020-01105-9}{main paper} \cite{rafiSemanticCognitiveTools2020} is 6.83, while $\nicefrac{1}{b}$ is only 2.46. Sellke at al. \cite{sellkeCalibrationValuesTesting2001} also discuss how $\frac{1}{1+\frac{1}{b}}$ is the Type-1 error rate for a particular type of conditional decision rule. \\ \indent
Grünwald et al. \cite{grunwaldSafeTesting2019} introduce a general concept they call an S-test statistic (where “S” stands for “safe”) for \textbf{H} given \textbf{A}, defined as any random variable \textit{S} satisfying $E_{M}(S)$ $\leq$ 1 under any model \textbf{M} obeying \textbf{H} and \textbf{A}. They also call this \textit{S} an “\textit{S}-value”. As noted above, our binary \textit{S}-value $S_{2}$ = $−\log_{2}(P)$ can be redefined using natural logs and thus rescaled to units of nats instead of units of bits, via $S_{e}$ =  $−\ln(P)$ = $−\log_{2}(P)\ln(2)$. $S_{e}$ is then an example of their S-value, since $E_{M}(S_{e})$ $\leq$ 1 when the random \textit{P}-value \textit{P} is valid or conservatively valid (uniform or dominated by a uniform random variable under \textbf{M}); it is also an example of a betting score \cite{shaferLanguageBettingStrategy2020} (hence “S” can also be taken as “information score”). Grünwald et al. \cite{grunwaldSafeTesting2019} discuss other \textit{S}-values, including those based on Bayes factors. \\ \indent
Finally, consider a 1-dimensional continuous parameter $\mu$ with test hypotheses \textbf{H} of the form $\mu\leq\mu_{0}$ and a specified alternative $\mu\geq\mu_{1}$ (or $\mu$ = $\mu_{0}$ with alternative $\mu$ = $\mu_{1}$) where $\mu_{0}$ < $\mu_{1}$. In this context, yet another S-word, “severity”, has been used to refer to the \textit{P}-value $p(\mu\geq\mu_{1})$ for $\mu\geq\mu_{1}$ (the lower tail of the test statistic \textit{m} − $\mu_{1}$ for a 1-sided test of $\mu$ = $\mu_{1}$ when using the estimate \textit{m} of $\mu$), which decreases as $\mu_{1}$ increases; see p. 345 and Fig, 5.5 of Mayo \cite{mayoStatisticalInferenceSevere2018}. Since the complement $p(\mu\leq\mu_{1})$ = 1 − $p(\mu\geq\mu_{0})$ is the \textit{P}-value for $\mu\leq\mu_{1}$, we find that (whatever the base) the corresponding \textit{S}-value function $s(\mu\leq\mu_{1})$ = $−\log(p(\mu\leq\mu_{1})$ measuring the information against $\mu\leq\mu_{1}$ increases as $p(\mu\geq\mu_{1})$ increases; thus $p(\mu\geq\mu_{1})$ varies directly with the information $s(\mu\leq\mu_{1})$ against $\mu\leq\mu_{1}$ (the case with alternative $\mu_{1}$ < $\mu_{0}$ is handled symmetrically). This so-called “severity” of the test of the original \textbf{H} ($\mu\leq\mu_{0}$) is not in fact a function of $\mu_{0}$ and so is identical for all $\mu_{0}$. Furthermore, it incorporates no information about background assumptions (e.g., whether treatment assignment was blinded) which bear heavily on practical notions of severity. We thus conclude that it is misleading to label $p(\mu\geq\mu_{1})$ as a severity measure, and it instead should be recognized and treated as the \textit{P}-value function it is.}}

\end{adjustwidth*}

\newpage
\begin{multicols}{2}
\small
\bibliographystyle{naturemag}
\bibliography{References.bib}
\end{multicols}
\end{document}